# Global Structure and Dynamics of Advection-Dominated Accretion Flows Around Black Holes

Ramesh Narayan

Harvard-Smithsonian Center for Astrophysics
60 Garden Street, Cambridge, MA 02138, and

Institute for Theoretical Physics
University of California, Santa Barbara, CA 93106

Shoji Kato and Fumio Honma

Department of Astronomy, Kyoto University
Sakyo-ku, Kyoto 606-01, Japan

**Abstract**

We present global solutions that describe advection-dominated accretion flows around black holes. The solutions are obtained by numerically solving a set of coupled ordinary differential equations corresponding to a steady axisymmetric height-integrated flow. The solutions satisfy consistent boundary conditions at both ends. On the inside, the flow passes through a sonic point and falls supersonically into the black hole with a zero-torque condition at the horizon. On the outside, the flow attaches to a normal thin accretion disk.

We obtain consistent transonic solutions for a wide range of values of the viscosity parameter $\alpha$, from 0.001 to 0.3. We do not find any need for shocks in our solutions, and disagree with previous claims that viscous accretion flows with low values of $\alpha$ must have shocks.

We compare the exact global solutions of this paper with a local self-similar solution which has been studied in the past. Although the self-similar solution makes significant errors close to the boundaries, we find that it nevertheless provides a reasonable description of the overall properties of the flow. We also compare two different forms of viscosity; one is based on a diffusion prescription while the other takes the shear stress to be simply proportional to the pressure. The results with the two prescriptions are similar.

We see a qualitative difference between solutions with low values of the viscosity parameter, $\alpha \lesssim 0.01$, and those with large values, $\alpha \gtrsim 0.01$. The solutions with low $\alpha$ have their sonic transitions occurring close to the radius of the marginally bound orbit. These flows are characterized by regions of super-Keplerian rotation, and have pressure maxima outside the sonic point. The solutions are similar in many respects to the hydrostatic thick tori developed previously as models of active galactic nuclei. In contrast, the solutions with large $\alpha$ have sonic transitions farther out, close to or beyond the marginally stable orbit, and have no super-Keplerian rotation or pressure maxima. We believe these flows will be nearly quasi-spherical down to the sonic radius and will not have empty funnels along the rotation axis. The large $\alpha$ solutions are more likely to be representative of real systems since most observations of advection-dominated flows in astrophysical sources indicate values of $\alpha \gtrsim 0.1$.

Subject headings: accretion, accretion disks: black hole physics: hydrodynamics, X-ray binaries, active galactic nuclei

# 1. Introduction

In the study of accretion disks, the model of thin disks developed by Shakura & Sunyaev (1973) and Novikov & Thorne (1973) has played a major role (see Pringle 1981 and Frank, King & Raine 1992 for reviews). This is a local model in the sense that the interactions between neighboring radial annuli are neglected. Consequently, the properties of the accreting gas at each radius are obtained by solving local algebraic equations, with some additional correction factors to allow for the effect of the inner boundary.

Paczyński & Bisnovatyi-Kogan (1981) and Muchotrzeb & Paczyński (1982) initiated the study of global models of thin accretion disks by writing down a set of coupled radial differential equations which include effects such as the radial pressure gradient and radial energy transfer. These equations with some modifications have been used by a number of workers to study transonic accretion flows around black holes (Muchotrzeb 1983, Matsumoto et al. 1984, Abramowicz et al. 1988) as well as possible shocks (Fukue 1987).

The thin accretion disk model describes a cooling-dominated flow where the viscous heating of the gas is balanced by local radiative cooling. Another regime of accretion is possible, corresponding to so-called advection-dominated flows, where the radiative cooling is very inefficient and most of the dissipated energy is advected into the black hole. Advection-dominated accretion can occur in two regimes:

1. At very high mass accretion rates, the optical depth becomes very high and the radiation can be trapped in the inflowing gas and be unable to diffuse out within the accretion time (Begelman 1978, Begelman & Meier 1982). A detailed analysis of these kinds of solutions was made by Abramowicz et al. (1988) who showed that, despite being radiation-pressure-dominated, such flows are stable to the Lightman & Eardley (1974) viscous instability as well as to the thermal instability (but see Kato, Abramowicz & Chen 1996).

2. The second regime is when the accretion rate is very small and the optical depth is very low. Early work on this kind of accretion was initiated by Rees et al. (1982). Recently, Narayan & Yi (1994, 1995a, b), Abramowicz et al. (1995) and Chen (1995) studied the dynamical properties of the flows, with special attention to the role of advection. In this regime, the low density of the accreting gas causes the cooling processes to become inefficient so that the radiative timescale is much longer than the accretion time. The net effect is the same as in the first case, namely that most of the dissipated energy becomes internal energy of the gas and is lost into the black hole.

The topological relationships among the above three branches of solutions, namely the thin disk solution, the optically-thick advection-dominated solution, and the optically-thin advection-dominated solution, as well as a fourth thermally unstable branch discovered by Shapiro, Lightman & Eardley (1976), have been worked out by Chen et al. (1995).

The optically-thin advection-dominated solution has been applied to several sources. Using a version of this model which involves a two-temperature plasma (Shapiro, Lightman & Eardley 1976, Rees et al. 1982, Narayan & Yi 1995b), models have been proposed for Sagittarius A* at the center of our Galaxy (Rees 1982, Narayan, Yi & Mahadevan 1995) and the supermassive black hole in the nucleus of NGC 4258 (Lasota et al. 1996), while Fabian & Rees (1995) and Mahadevan (1996) have suggested that many nearby elliptical galaxies may have large black holes in their nuclei accreting in an advection-dominated mode. In addition, Narayan, McClintock & Yi (1996) have developed successful models of the soft X-ray transients A0620-00, V404 Cyg and Nova Muscae 1991 in their quiescent states, and Narayan (1996) has shown that this branch of solutions may be applicable also to some high-luminosity systems. All of these applications have been based on a local self-similar solution of advection-dominated accretion (Narayan & Yi 1994, Spruit et al. 1987). This solution is expected to be a good description of the flow away from the boundaries. However, the spectral calculations assume that the local solution is valid all the way down to the marginally stable orbit. No estimate of the error due to this approximation is currently available since it requires solving the global problem with boundary conditions.

Very few global studies of advection-dominated accretion exist in the literature. Matsumoto, Kato and Fukue (1985) published global solutions of optically-thin two-temperature advection-dominated flows, while Abramowicz et al. (1988) (see also Honma, Matsumoto & Kato 1991, Chen & Taam 1993) presented results for the optically-thick branch of solutions, and Igumenshchev, Chen & Abramowicz (1996) described some numerical simulations of advection-dominated flows around black holes. We present in this paper a detailed numerical study of the properties of global solutions, concentrating on the nature of the flow near the black hole, the dependence of the properties on the value of the viscosity parameter $\alpha$, and comparing the global and local solutions.

One of the features of advection-dominated flows is that the pressure in the accreting gas becomes very high, approaching the virial pressure. Consequently, the gas does not have a disk-like morphology but a more nearly spherical form (Narayan & Yi 1995a), at least at large radii from the black hole. What is the morphology close to the black hole?

A great deal of work has been done on non-viscous, non-accreting hydrostatic fluid configurations around black holes (Fishbone & Moncrief 1976, Abamowicz, Jaroszyński & Sikora 1978, Kozlowski, Jaroszyński & Abramowicz 1978, Paczyński & Wiita 1980). These models are referred to as "thick disks" or "thick tori." Generic features of the models include (i) the existence of a radial pressure maximum in the accreting gas, (ii) super-Keplerian rotation over a range of radius inside the pressure maximum, and (iii) the presence of an empty funnel around the rotation axis. What is the relationship between these thick hydrostatic tori and global advection-dominated accretion flows? In particular, do advection-dominated flows have pressure maxima, super-Keplerian rotation, and empty funnels? We attempt to provide some answers to these questions.

In §2 we set up the differential equations and boundary conditions of the global problem, and in §3 we describe the results. The paper concludes with a discussion in §4.

## 2. Theory

### 2.1. Dynamical Equations

We consider a steady state axisymmetric accretion flow, employing a height-integrated set of equations, similar to the "Slim Disk Equations" of Abramowicz et al. (1988). In this formulation, all physical variables are functions only of the cylindrical radius $R$. The continuity equation takes the form

$$\frac{d}{dR}(2\pi R \cdot 2H \cdot \rho v) = 0, \qquad (2.1)$$

where $\rho$ is the density of the gas, $H$ is the vertical "half-thickness," and $v$ is the radial velocity. Equation (2.1) integrates to give the mass accretion rate,

$$-4\pi R H \rho v = \dot{M} = \text{constant}. \qquad (2.2)$$

We define $v$ to be negative when the flow is inward.

In the spirit of thin accretion disk theory, we express the vertical thickness of the disk in terms of the isothermal sound speed $c_s$ of the gas,

$$H = \left(\frac{5}{2}\right)^{1/2} \frac{c_s}{\Omega_K}, \qquad (2.3)$$

where $\Omega_K$ is the Keplerian angular velocity. The coefficient $(5/2)^{1/2}$ in this equation is chosen on the basis of the self-similar solution obtained by Narayan & Yi (1995a), which shows that a perfectly spherically-symmetric flow has $c_s^2 = (2/5)\Omega_K^2 R^2$. If we set $c_s^2 = (2/5)\Omega_K^2 R^2$ in equation (2.3), we obtain $H = R$, and equation (2.2) then gives the correct result for a spherical flow, namely $\dot{M} = -4\pi R^2 \rho v$.

Advection-dominated flows tend to be nearly spherical, and one may wonder if height-integrated equations can be used to describe such flows. This question has been considered by Narayan & Yi (1995a), who showed that, at least in the self-similar regime which operates far from boundaries, the agreement between the approximate height-integrated solution and the exact solution is remarkably good. The interesting point is that height-integration must be interpreted, not as a vertical integration over cylindrical coordinate $z$, but rather as a spherical integration over polar angle $\theta$. With this interpretation the height-integrated equations are a very good representation of the behavior of both thin disks and nearly-spherical flows. This result, however, is strictly true only if $H/R$ is independent of radius, as in the self-similar solutions. For more general flows, one would expect correction terms of order

$$e = \frac{\partial \ln(H/R)}{\partial \ln R}. \qquad (2.4)$$

Our dynamical equations effectively assume $e = 0$, though the energy equation (2.12) below includes the effect of $H/R$ variations in the definition of the entropy. It is difficult to assess the magnitude of the error we make by ignoring terms of order $e$, but we note that the solutions presented in this paper do have a small value of $e$ over a wide range of radius (see Fig. 7).

We assume that the gravitational potential of the central black hole is described by the potential introduced by Paczyński & Wiita (1980),

$$\Phi(R) = -\frac{GM}{(R - R_g)}, \qquad R_g \equiv \frac{2GM}{c^2}, \qquad (2.5)$$

where $M$ is the mass of the black hole and $R_g$ is its gravitational radius. The Keplerian angular velocity takes the form

$$\Omega_K^2 = \frac{GM}{(R - R_g)^2 R}. \qquad (2.6)$$

The steady state radial momentum equation of the accreting gas is then given by

$$v\frac{dv}{dR} = -\Omega_K^2 R + \Omega^2 R - \frac{1}{\rho}\frac{d}{dR}(\rho c_s^2), \qquad (2.7)$$

where $\Omega$ is the angular velocity of the gas. The last term is the acceleration due to the pressure gradient, where we have written the pressure in terms of the isothermal sound speed,

$$p = \rho c_s^2. \tag{2.8}$$

The steady state angular momentum equation in the presence of viscosity takes the form

$$v \frac{d}{dR}(\Omega R^2) = \frac{1}{\rho R H} \frac{d}{dR}\left(\nu \rho R^3 H \frac{d\Omega}{dR}\right), \tag{2.9}$$

where $\nu$ is the kinematic coefficient of viscosity. In the spirit of Shakura & Sunyaev (1973), we write $\nu$ as

$$\nu = \alpha \frac{c_s^2}{\Omega_K}, \tag{2.10}$$

where we assume $\alpha$ to be a constant, independent of $R$. Substituting equation (2.10) in (2.9) and integrating, we obtain

$$\frac{d\Omega}{dR} = \frac{v \Omega_K (\Omega R^2 - j)}{\alpha R^2 c_s^2}. \tag{2.11}$$

The integration constant $j$ represents the specific angular momentum per unit mass accreted by the black hole, and is an eigenvalue of the problem (cf. Abramowicz et al. 1988, Popham & Narayan 1991, 1992, Paczyński 1991). We determine $j$ self-consistently through physical boundary conditions.

Finally, we consider the energy equation. We write this as an entropy equation by setting the Lagrangian rate of change of the entropy of a parcel of gas equal to the rate of heating by viscous dissipation minus the rate of radiative cooling. For the most part, we are interested in advection-dominated flows where the radiative cooling is negligibly small. However, in a few applications we do consider the possibility that the cooling may be important. We therefore write the energy equation in a general form, where we take the cooling to be a factor $(1 - f)$ times the heating rate, with $f$ being a function of $R$. The energy equation then takes the form

$$\frac{\rho v}{(\gamma - 1)} \frac{dc_s^2}{dR} - c_s^2 v \frac{d\rho}{dR} = \frac{f \alpha \rho c_s^2 R^2}{\Omega_K} \left(\frac{d\Omega}{dR}\right)^2, \tag{2.12}$$

where $\gamma$ is the ratio of specific heats of the accreting gas. An advection-dominated flow has $f(R) = 1$, while a cooling-dominated flow has $f(R) \ll 1$.

We concentrate on advection-dominated flows in this paper and set $f(R) = 1$. This is an excellent approximation whenever the radiative cooling is less than a few per cent of the heating rate. Most applications considered so far (Narayan et al. 1995, 1996, Fabian & Rees 1995, Lasota et al. 1996) are comfortably in this regime. When the cooling is very inefficient, it is not necessary to include all the details of the cooling while calculating the dynamics. Rather, one can consider the cooling to be a minor perturbation and solve for the dynamics under the assumption of zero cooling. After solving for the dynamics, one can go back and compute the actual energy lost via cooling and calculate the spectrum of the emission.

## 2.2 Boundary Conditions

The dynamical equations described in §2.1 consist of three first-order differential equations, namely (2.7), (2.11) and (2.12), each of which requires a boundary condition. In addition, the eigenvalue $j$ is an unknown which needs to be determined self-consistently. Finally, since we are interested in transonic solutions which go through a sonic point, the sonic radius $R_s$ is yet another unknown which needs to be determined self-consistently. Thus, the problem requires a total of five boundary conditions.

We solve the equations numerically on a radial grid which goes out to a maximum radius $R_{out} = 10^6 R_s$. At $R = R_{out}$ we assume that the gas starts off in a state which corresponds to a thin accretion disk. This allows us to apply two boundary conditions at $R = R_{out}$:

$$\Omega = \Omega_K, \qquad R = R_{out}, \tag{2.13}$$

$$c_s = 10^{-3} \Omega_K R, \qquad R = R_{out}. \tag{2.14}$$

Combining equations (2.3) and (2.14) we see that the vertical thickness at $R = R_{out}$ is $H \sim 10^{-3} R$.

Since the accretion flow makes a sonic transition on the inside and flows supersonically into the black hole, the equations become singular at the sonic radius. The singularity provides two boundary conditions at $R = R_s$. To derive these conditions, let us substitute for $d\Omega/dR$ in equation (2.12) using equation (2.11), and substitute for $\rho$ using equation (2.2). This gives

$$\frac{(\gamma+1)}{(\gamma-1)} \frac{d\ln c_s}{dR} = -\frac{d\ln|v|}{dR} + \frac{d\ln \Omega_K}{dR} - \frac{1}{R} + \frac{f\Omega_K v(\Omega R^2 - j)^2}{\alpha R^2 c_s^4}. \tag{2.15}$$

Using this equation to eliminate $dc_s/dR$ in the radial momentum equation (2.7), we then obtain the following differential equation for $dv/dR$:

$$\left[\frac{2\gamma}{(\gamma+1)} - \frac{v^2}{c_s^2}\right] \frac{d\ln|v|}{dR} = \frac{(\Omega_K^2 - \Omega^2)R}{c_s^2} - \frac{2\gamma}{(\gamma+1)} \left(\frac{1}{R} - \frac{d\ln\Omega_K}{dR}\right) + \frac{f(\gamma-1)\Omega_K v(\Omega R^2 - j)^2}{\alpha(\gamma+1)R^2 c_s^4}. \tag{2.16}$$

Since $dv/dR$ must be well-behaved across the sonic point, this gives two boundary conditions at $R = R_s$:

$$v^2 - \frac{2\gamma}{(\gamma+1)} c_s^2 = 0, \qquad R = R_s, \tag{2.17}$$

$$\frac{(\Omega_K^2 - \Omega^2)R}{c_s^2} - \frac{2\gamma}{(\gamma+1)} \left(\frac{1}{R} - \frac{d\ln\Omega_K}{dR}\right) + \frac{f(\gamma-1)\Omega_K v(\Omega R^2 - j)^2}{\alpha(\gamma+1)R^2 c_s^4} = 0, \qquad R = R_s. \tag{2.18}$$

These two conditions automatically ensure that the flow will make a smooth sonic transition at $R = R_s$.

The final boundary condition is obtained from the angular momentum equation. Because the viscous stress term in equation (2.9) is diffusive in form, signals due to the shear stress propagate at infinite speed (cf. Narayan 1992) and can propagate backwards even in the supersonic zone inside $R = R_s$. Therefore, the fifth boundary condition has to be applied at the black hole horizon. Since the horizon cannot support a shear stress, we impose a no-torque condition ($d\Omega/dR = 0$) at $R = R_g$, which corresponds to the boundary condition

$$\Omega R^2 - j = 0, \qquad R = R_g. \tag{2.19}$$

This condition states that the rate at which angular momentum falls into the black hole is equal to the mass accretion rate multiplied by the specific angular momentum of the accreting material at the horizon. The condition is intuitively obvious and has been used in previous investigations of accretion flows around black holes (e.g. see the detailed discussion of Abramowicz et al. 1988 and references therein). As already mentioned, the boundary condition arises because the diffusion operator in the angular momentum equation permits viscous signals to propagate backward in the supersonic zone. This makes it necessary to impose a downstream boundary condition on the angular momentum. Although technically the no-torque condition must be applied at the horizon, in practice, the condition could equally well be applied at any other radius between the horizon and the sonic point, or even at the sonic radius itself. The results change only slightly in quantitative details when this is done.

A more rigorous approach to this problem is to replace the diffusive model of viscosity by a causality-enforcing formalism of viscosity as described by Narayan, Loeb & Kumar (1994) or Kato (1994). Such a formalism will automatically ensure that signals do not propagate backward in the supersonic zone. Therefore, the condition (2.19) at the horizon will be replaced by other physical boundary conditions on the stress tensor at the sonic radius. We have not yet calculated self-consistent models with this approach, but anticipate that the results will be qualitatively similar to those described in this paper.

Equations (2.13), (2.14), (2.17), (2.18) and (2.19) provide the five boundary conditions needed to solve the problem. Since these conditions are applied at three different radii, viz. $R_{out}$, $R_s$ and $R_g$, we have a fairly challenging numerical boundary value problem. For given choices of $\alpha$, $\gamma$ and $f(R)$, we solve the equations as follows. We assume a value for the sonic radius $R_s$ and solve the differential equations on a grid between $R = R_s$ and $R = R_{out}$ and optimize the value of $j$ at the same time using a relaxation method (Press et al. 1986). We then integrate in from $R = R_s$ using equations (2.17) and (2.18) to start off the integration at the critical point. For a wrong choice of $R_s$, the solution will not satisfy the condition (2.19) but will diverge. We therefore vary the value of $R_s$ until the fifth condition is also satisfied. We then have a self-consistent and unique global solution to the problem. The solution will of course have a well-behaved sonic transition at $R_s$ because of the explicit introduction of the boundary conditions (2.17), (2.18).

Of the five boundary conditions, those given in equations (2.17), (2.18) and (2.19) are quite important (though as we have said 2.19 could be applied anywhere between the horizon and the sonic radius), while the two conditions at the outer edge, equations (2.13) and (2.14), are relatively less critical. Thus, even if we change the values of $\Omega$ and $c_s$ at the outer edge by fairly large factors, we find that the solution is modified only in the vicinity of $R_{out}$, while the solution at smaller radii is essentially unaffected (see Fig. 5 below). This point was also recognized by Abramowicz et al. (1988).

It should be emphasized that in our approach the sonic radius, $R_s$, and the specific angular momentum accreted by the black hole, $j$, are determined self-consistently via physically motivated boundary conditions. This is an improvement over other approaches in the literature where these quantities appear to be selected arbitrarily (e.g. Chakrabarti & Titarchuk 1995, see the discussion in their Appendix).

### 2.3 Self-Similar Solution

The aim of this paper is to obtain numerical global solutions to the problem posed above. One of our interests is to compare the global solutions to a self-similar solution discovered by Spruit et al. (1987) and Narayan & Yi (1994). This particular solution is obtained by assuming that $R \gg R_g$, so that the potential (2.5) simplifies to the Newtonian form. If we assume that the flow variables have power-law dependences on the radius and take $f$ to be independent of $R$ (a natural assumption for an advection-dominated flow where $f \approx 1$ at all $R$), then the equations permit the following special solution:

$$v = -\frac{(5+2\epsilon')g(\alpha,\epsilon')}{3\alpha}\Omega_K R, \tag{2.20}$$

$$c_s = \left[\frac{2(5+2\epsilon')g(\alpha,\epsilon')}{9\alpha^2}\right]^{1/2}\Omega_K R, \tag{2.21}$$

$$\Omega = \left[\frac{2\epsilon'(5+2\epsilon')g(\alpha,\epsilon')}{9\alpha^2}\right]^{1/2}\Omega_K, \tag{2.22}$$

where

$$g(\alpha,\epsilon') = \left[1 + \frac{18\alpha^2}{(5+2\epsilon')^2}\right]^{1/2} - 1, \qquad \epsilon' = \left(\frac{1}{f}\right)\frac{(5/3-\gamma)}{(\gamma-1)}. \tag{2.23}$$

Narayan & Yi (1994) speculated, on the basis of some numerical experiments, that the self-similar solution is the "natural" state for an advection-dominated flow. They suggested that, even if there are boundary conditions which do not match the self-similar solution, the accreting gas would tend toward the self-similar form away from the boundaries.

## 3. Results

### 3.1 Two Classes of Global Solutions

Figures 1 and 2 show the variations with $R$ of the radial velocity $v$, sound speed $c_s$, and angular momentum $l = \Omega R^2$ for a series of advection-dominated solutions with $\alpha = 0.001$, 0.003, 0.01, 0.03, 0.1, and 0.3, respectively. In these solutions, $\gamma = 1.5$, as appropriate for an optically-thin advection-dominated flow with roughly equal amounts of gas and magnetic pressure (Narayan & Yi 1995b). Also, we have set the advection parameter $f$ equal to unity at all $R$.

At the outer edge, all the solutions have very low values of $v$ and $c_s$, and the rotation is Keplerian, $\Omega = \Omega_K$, according to the boundary conditions (2.13), (2.14) which correspond to a very thin accretion disk. Within a short distance from the outer boundary, however, we see that the solutions are modified significantly. The sound speed increases by a large amount until it is approximately equal to the local virial speed: $c_s \sim \Omega_K R$. This is, of course, to be expected since there is no cooling and therefore all the gravitational binding energy released as the gas flows in is converted into internal thermal energy of the gas.

As the sound speed goes up, the viscosity increases (see eq. 2.10) and this causes the radial velocity of the accretion to increase. Roughly, we have $v \sim \nu/R \sim \alpha\Omega_K R$. Thus, the accretion velocity becomes of order $\alpha$ times the free-fall velocity. This corresponds to quite rapid accretion, especially for large values of $\alpha \sim 0.1 - 0.3$. Note, however, that the radial velocity is subsonic for all values of $\alpha$.

The large radial velocity causes the density to drop (see eq 2.2) compared to its value in a thin accretion disk. For many optically-thin cooling processes (e.g. bremsstrahlung and even synchrotron if the magnetic field strength is in equipartition with the gas pressure) this means that the cooling rate per unit mass reduces. This drives the gas toward advection-domination, so that the assumption $f \to 1$ is self-consistent. The details of the cooling are outside the scope of this paper; as we argued earlier, when the cooling rate is very much less than the heating rate, as it is in an advection-dominated flow, it is appropriate to neglect the cooling for the purposes of calculating the dynamics. The reader is referred to Narayan & Yi (1995b) and Abramowicz et al. (1995) for a discussion of cooling.

Another consequence of the increasing sound speed is that the pressure gradient becomes important and the gas achieves partial pressure-support against the gravity of the black hole. As a result, the gas rotates with a sub-Keplerian angular velocity.

The above characteristics are maintained over a range of radii, but as the gas approaches the gravitational radius, a series of changes takes place. These changes are driven by the presence of the sonic transition at $R = R_s$. The radial velocity has to increase from its "cruising" value $v \sim \alpha c_s$ to $v \sim c_s$ at the sonic point (see eq. 2.17). This change is most dramatic for the low $\alpha$ solutions where the cruising $v \ll c_s$. The increase in $v$ causes a corresponding decrease in $\rho$, leading to a reduction in the pressure support. The rotation $\Omega$ therefore increases, approaching the Keplerian rotation (Fig. 2).

While these characteristics are common to all the solutions, we confirm the finding of Matsumoto et al. (1985) that the solutions with low values of $\alpha$ have a qualitatively different behavior at small $R$ than those with large $\alpha$. There are three related features in which the two classes of solution differ:

(i) Solutions with low values of $\alpha$ have their sonic transition near the marginally bound radius, $R \sim 2R_g$. For instance, the solutions with $\alpha = 0.001$, $0.003$ have $R_s = 2.096R_g$, $2.152R_g$ respectively. However, as $\alpha$ increases the sonic radius moves out significantly. For instance, for $\alpha = 0.1$, we have $R_s = 3.066R_g$, which is outside the last stable orbit $(3R_g)$, while for $\alpha = 0.3$, the sonic point is even farther out, $R_s = 5.315R_g$. The moving out of the sonic radius with increasing $\alpha$ was noted by Abramowicz et al. (1988).

(ii) For low values of $\alpha$, the rotation is super-Keplerian over a range of radii just outside the sonic point. This is shown in Fig. 3, which gives an expanded view of the inner region of Fig. 2. The solutions with large values of $\alpha$, on the other hand, are sub-Keplerian all the way down to the black hole. Especially in the cases when $\alpha = 0.1$ and $0.3$, we see that the rotation is substantially sub-Keplerian even near the last stable orbit, $R = 3R_g$.

(iii) Figure 4 shows the pressure in the inner regions of the various solutions. At low values of $\alpha$, we see that the pressure goes through a maximum outside the sonic radius and the pressure falls inwards as the gas crosses the sonic point. For large $\alpha$, on the other hand, the pressure increases monotonically inward. The difference is easy to understand. Two competing effects influence the pressure: $p$ increases inward because of the (spherically) converging nature of the flow, but decreases because of the radial acceleration which tends to decrease the density. In low $\alpha$ flows, the acceleration is enormous as $v$ rises toward the sonic point (see Fig. 1), and this causes $p$ to fall. But in flows with large values of $\alpha$, there is only modest acceleration and the radial convergence dominates.

The boundary between the two classes of solutions seems to occur roughly at $\alpha \sim 0.01$. We note that Muchotrzeb (1983), Matsumoto et al. (1984), and Muchotrzeb-Czerny (1986) found a change in the behavior of global thin accretion disk flows at around $\alpha \sim 0.02 - 0.05$.

We interpret the change in the character of the flow between small and large $\alpha$ as due to differences in the importance of viscosity. At low $\alpha$, viscosity is weak, and the accreting gas is in rough hydrostatic equilibrium until it reaches very close to the sonic radius. In this regime, the studies of hydrostatic tori (e.g. Fishbone & Moncrief 1976, Abramowicz et al. 1978, Kozlowski et al. 1978, Paczyński & Wiita 1981) are relevant and we expect our solutions to behave like these thick disk models. Key features of thick disk models are that (i) the inner edge is close to the marginally bound orbit, (ii) the rotation is super-Keplerian close to the sonic radius, and (iii) the gas has a pressure maximum at the edge of the super-Keplerian zone. Our low-$\alpha$ solutions show all of these features. The only quantitative difference is that the pressure maximum occurs at a fairly small radius and so our tori are not as large as some models described in the literature (e.g. Fishbone & Moncrief 1976, Paczyński & Wiita 1980).

At large values of $\alpha$, however, viscosity plays an important role in the dynamics, so that the radial momentum equation differs from simple hydrostatic equilibrium. The important new term is the dynamic pressure gradient, $v(dv/dR)$, which becomes large because $v$ is large in these solutions. The gas experiences significant dynamical acceleration, and as a result the usual properties of thick disk models are not seen. Instead, the gas continues to accrete in a sub-Keplerian fashion all the way through the sonic point. Also, the sonic point moves out because of the slow rotation. It is known that gas with Keplerian angular momentum cannot reach the black hole via free-fall from any radius $R > 3R_g$. However, when the rotation is sub-Keplerian the gas can fall into the black hole even from $R > 3R_g$.

Our physical interpretation of these results is that in flows corresponding to low values of $\alpha$, the gas is pushed across the sonic point by the *pressure gradient*. This obviously requires a pressure maximum outside the sonic radius. In contrast, the gas in flows with large $\alpha$ is pushed across the sonic point by the action of *viscosity* which removes angular momentum efficiently and causes the gas to fall in. This interpretation of the change of global structure as a function of $\alpha$ was already recognized by Matsumoto et al. (1984) in the case of geometrically-thin, optically-thick disks. The transition between the two regimes happens continuously as $\alpha$ is varied. Nevertheless, we believe that the physics in the two limits is qualitatively very distinct.

The fact that the pressure increases monotonically inward in the large $\alpha$ flows suggests to us that these flows will not be toroidal in morphology as in the usual models of thick tori. Instead, we suggest that the flow remains *quasi-spherical* all the way down to the sonic radius. Similarly, the fact that the flow does not become super-Keplerian at any radius suggests that there will not be empty funnels along the rotation axis. Recall that in thick disk models the funnel walls are pushed out from the rotation axis by the centrifugal action of gas rotating with super-Keplerian velocities. We do not have such a large outward acceleration in our flows. We thus believe that advection-dominated flows with large values of $\alpha$ are qualitatively very different from standard models of thick tori.

One point we wish to emphasize is that we obtain well-behaved transonic solutions for all values of $\alpha$ from $\alpha \to 0$ to $\alpha \gtrsim 0.3$, and we find no necessity for shocks at any reasonable value of $\alpha$. We obtain shock-free transonic solutions in two quite independent calculations, namely those described in this section and those discussed in §3.3. In addition, the same result is also obtained in a third independent calculation by Chen, Abramowicz & Lasota (1996). Further, the result is essentially independent of $\gamma$ since we have obtained shock-free transonic solutions for all $\gamma$ ranging from $\gamma = 4/3$ to $5/3$, in each case with $\alpha$ ranging from $0.001$ to $> 0.1$.

Chakrabarti & Titarchuk (1995) state on the basis of a number of earlier papers by Chakrabarti that viscous flows with low values of $\alpha$ have shocks while those with large values of $\alpha$ do not. We do not agree with their result. Chakrabarti & Titarchuk (1995) further claim that low-$\alpha$ flows have sub-Keplerian rotation at large radii while high-$\alpha$ flows are Keplerian at all radii except very close to the black hole. Again we disagree. As Fig. 2 shows, we find that all advection-dominated flows have sub-Keplerian rotation at large radii, regardless of the value of $\alpha$. At small radii, we find that low-$\alpha$ flows go super-Keplerian near the marginally stable orbit ($R \sim 3R_g$) while large-$\alpha$ flows remain substantially sub-Keplerian at all radii. These results are confirmed in all three independent calculations mentioned in the previous paragraph.

### 3.2 Comparison with the Self-Similar Solution

The dotted lines in Figs. 1 and 2 show the self-similar solution of Narayan & Yi (1994) for $v$, $c_s$ and $\Omega$ (eqs. 2.20–2.22). We see that the exact global solutions achieve approximate self-similar behavior within a short distance from the outer boundary. Considering that $v$ and $c_s$ in particular have values at the outer boundary which are orders of magnitude different from the self-similar solution, the approach to self-similarity is quite impressive. The rotation profile approaches self-similarity more slowly. The self-similar rotation rate is almost a factor of 3 below Keplerian. The rotation profile has to negotiate this jump while at the same time maintaining $d\Omega/dR < 0$ in order to transfer angular momentum outward. The latter requirement moderates the approach to self-similarity.

Figure 5 shows three solutions which differ in the choice of the outer boundary conditions. The plot confirms that the influence of the outer conditions dies down as the flow moves in. Indeed, the three solutions are so similar on the inside that the sonic radius $R_s$ and the angular momentum eigenvalue $j$ differ by only 2 parts in $10^5$ among the three solutions.

At small radii, two effects cause the global solutions to deviate from the self-similar form. First, the potential (2.5) differs from the pure Newtonian form which was assumed to derive equations (3.1)-(3.4). This effect becomes significant for $R \lesssim 10R_g$. More importantly, the radial Mach number of the flow has to increase from its self-similar value, which is $\sim \alpha$, to unity as $R$ approaches $R_s$. This induces deviations from self-similarity which are especially large when $\alpha$ is small. Note, however, that the sound speed is very close to the self-similar form at nearly all radii in all solutions.

The solutions discussed so far in the paper correspond to pure advection-dominated flows with $f = 1$ at all $R$. How does the flow behave when there is cooling and the parameter $f$ varies smoothly with radius? Figure 6 shows an example where we assume $f = 1$ for $\log(R/R_s) < 2.5$, $f = f_{out} = 10^{-5}$ for $\log(R/R_s) > 5.5$, and we take $f$ to vary smoothly between the two limits as follows:

$$f(R) = f_{out} + \frac{(1 - f_{out})}{2}\left[1 + \cos\left[(\log(R/R_s) - 2.5)\pi/3\right]\right]. \qquad (3.6)$$

This model corresponds to a case where the accretion flow behaves like a standard thin disk at radii $\log(R/R_s) > 5.5$ and then makes a smooth transition to a fully advection-dominated flow. The functional form is selected arbitrarily, subject to the requirements of smoothness. We have chosen $\alpha = 0.1$ for this example.

We have solved the global problem for this model of $f(R)$, using as the outer boundary conditions the self-similar values of $v$ and $c_s$ (eqs. 2.20, 2.21) corresponding to $f = f_{out}$. Figure 6 shows the results. For comparison, we show by dashed lines the variations we would expect if the flow were to adjust instantly to the local self-similar solution at each $R$ with the local value of $f(R)$. The comparison suggests that the local model does not make a large error. Incidentally, a comparison of this solution with the corresponding solution in Fig. 1 for the same choice of $\alpha$ and $\gamma$ reveals that $R_s$ and $j$ vary by only $7.4 \times 10^{-4}$ and $8.8 \times 10^{-4}$ respectively. This again confirms that $R_s$ and $j$ are determined by physics on the inside and are quite insensitive to conditions on the outside.

The self-similar solution has an unexpected property (Narayan & Yi 1994, 1995a), namely the gas has a positive Bernoulli parameter,

$$Be = \frac{1}{2}v^2 + \frac{1}{2}\Omega^2 R^2 - \Phi(R) + \frac{\gamma}{(\gamma - 1)}c_s^2. \tag{3.7}$$

Since the Bernoulli parameter is conserved in inviscid adiabatic flows, the positivity of $Be$ suggests that the accreting gas may be capable of flowing out to infinity with a net positive (kinetic) energy. Narayan & Yi (1994, 1995a) speculated that this property may lead to a purely hydrodynamic mechanism to produce jets in accretion flows. However, a specific model has yet to be developed.

Figure 7 shows the variation of $Be$ with radius for the solutions shown in Figs. 1-4. We see that the accreting gas has a positive $Be$ over most of the flow, as expected from the self-similar solution. Indeed, $Be$ appears to be a little larger than the self-similar value; this is because, close to the black hole, the radial and tangential velocities are larger in the global solution than in the self-similar solution. For the particular flows shown in Fig. 7, the maximum outflow velocity we can expect at infinity is about $0.1c - 0.15c$. Narayan & Yi (1995a, see their Fig. 4) discuss the variation of $Be$ with polar angle $\theta$ in the self-similar regime.

Figure 7 also shows the error parameter $e$, defined in equation (2.4), for the various solutions. We note that $e$ is quite small over a wide range of radius, thus confirming the validity of our approximations at these radii. Near the boundaries, however, $e$ does become large in magnitude. It is not clear how much of an error this introduces in the solutions. An improved model which takes into account the dynamics of "vertical motions" is needed if one wishes to model the boundary regions more accurately.

### 3.3 Effect of Varying the Viscosity Prescription

The calculations described so far are based on a diffusion-type formula for the shear stress, using the kinematic viscosity coefficient given in equation (2.10). Shakura & Sunyaev (1973) originally wrote a simpler prescription for the shear stress of the form

$$\text{Shear Stress} = -\alpha p \frac{d\ln\Omega_K}{d\ln R}, \tag{3.7}$$

where $p$ is the pressure. With this prescription, equation (2.9) is modified to

$$v\frac{d}{dR}(\Omega R^2) = \frac{1}{\rho R H}\frac{d}{dR}\left(\alpha p R^2 H \frac{d\ln\Omega_K}{d\ln R}\right), \tag{3.8}$$

which integrates to give the following simpler relation instead of equation (2.11):

$$v\left(\Omega R^2 - j\right) = \alpha c_s^2 R \frac{d\ln\Omega_K}{d\ln R}. \tag{3.9}$$

In addition, the energy equation (2.12) is modified to

$$\frac{\rho v}{(\gamma-1)}\frac{dc_s^2}{dR} - c_s v \frac{d\rho}{dR} = f\alpha p \frac{d\ln\Omega_K}{d\ln R}\left(R\frac{d\Omega}{dR}\right). \tag{3.10}$$

We have calculated global solutions corresponding to these modified equations for a range of values of $\alpha$. Since the number of first-order differential equations is reduced by one, we need one fewer boundary condition. In the present case signals cannot propagate backward from the supersonic zone (Matsumoto et al. 1984). Hence the inner boundary condition (2.19) is unnecessary. We impose the outer boundary conditions (2.13) and (2.14) at $R = R_{out}$, where $R_{out} = 10^6 R_g$. Our procedure for vertical integration of the basic two-dimensional equations here is the same as in Matsumoto et al. (1984) and is slightly different from that used in the previous sections. Because of this, the sonic conditions are a little different from equations (2.17) and (2.18).

Figures 8 and 9 show the results for this viscosity prescription, and should be compared with the equivalent Figs. 1 and 3 for the diffusion prescription. We see that the radial distributions of the velocity, temperature and specific angular momentum for $R \gg R_s$ are almost the same in the two cases. A comparison of equations (2.11) and (3.9) shows that the radial velocity in the present case is larger than with the diffusion formula by a factor $\sim \Omega_K/\Omega$. This causes the sonic radius to shift outward, as can be seen by comparing Figs. 8 and 1. Despite this difference, the qualitative properties around $R \sim R_s$ are the same as before. That is, for low values of $\alpha$, we find $R_s \sim 2R_g$ and a region with super-Keplerian rotation around $R \sim 3R_g$ (see Fig. 9), whereas for large values of $\alpha$, we find $R_s > 3R_g$ and sub-Keplerian rotation all the way down to the black hole. Similarly, we find that the pressure has a maximum when $\alpha$ is small, but increases monotonically inward when $\alpha$ is large.

## 4. Summary and Discussion

In this paper we have calculated global solutions of advection-dominated accretion flows around black holes, paying careful attention to the transonic region close to the black hole. Although the gas in our solutions has a nearly spherical morphology, similar to that in Bondi (1952) spherical accretion, the dynamics is much more similar to that which occurs in *thin* accretion disks. The sonic radius $R_s$ in our solutions, for instance, occurs in the range $R_s \sim (2-5)R_g$. This is similar to $R_s = 3R_g$ in thin disks and $R_s \sim 2R_g$ in thick tori, but is very different from Bondi accretion where for $\gamma < 5/3$ we generally have $R_s \gg R_g$.

The reason for the difference lies in the fact that the accreting gas in our solutions has angular momentum which provides at least partial support against gravity. Only by losing angular momentum can the gas fall into the black hole. This means that the accretion has to be fundamentally regulated by the rate of viscous transport of angular momentum. Pure spherical accretion on the other hand involves nothing more than a simple competition between gravity and pressure, with neither angular momentum nor viscosity playing any role. Abramowicz & Zurek (1981) have shown that there is a clear mathematical distinction between spherical accretion and rotating disklike accretion in the limit of weak viscosity.

We have obtained global advection-dominated solutions for a wide range of values of the viscosity parameter $\alpha$, from $\alpha = 0.001$ to 0.3 (§2.1). The calculations show that there is a distinct difference between flows with low values of $\alpha < 0.01$ and those with large $\alpha > 0.01$.

Flows with low $\alpha$ have (i) $R_s \sim 2R_g$, (ii) super-Keplerian rotation over a range of radius just outside the sonic radius, and (iii) a pressure maximum outside the sonic radius. All of these features are characteristic of thick disks or thick tori which have been discussed in the past as possible models of active galactic nuclei (Fishbone & Moncrief 1976, Abamowicz et al. 1978, Kozlowski et al. 1978, Paczyński & Wiita 1980). We could therefore describe our low $\alpha$ solutions as improved versions of thick disk models. The principal improvement in our models is that we self-consistently take account of angular momentum and energy transfer, which was missing in the earlier work.

The solutions we obtain for large values of $\alpha$ are very different. They have (i) $R_s > 3R_g$ (for $\alpha \gtrsim 0.1$), (ii) sub-Keplerian rotation at all radii, and (iii) no pressure maximum outside the sonic point. The dynamics in these large $\alpha$ solutions is strongly dominated by viscosity, and the flows are physically very different from the thick tori mentioned above, which are essentially hydrostatic objects. The properties of large $\alpha$ flows were originally discussed by Matsumoto et al. (1984) and Chen & Taam (1993) in connection with optically-thick accretion disks.

The physical origin of the difference between low and high $\alpha$ flows has been discussed by Matsumoto et al. (1984, 1985). For low $\alpha$, the radial velocity of the gas (which is $\sim \alpha \Omega_K R$) is small and the structure of the flow is determined essentially by radial hydrostatic balance. This explains the close similarity of these flows to thick disk models since the latter are generally constructed under the hydrostatic assumption. In these cases, the accreting gas is pushed across the sonic point mostly by the action of the pressure in the torus. In contrast, for large $\alpha$, dynamical terms become important in the radial momentum equation and the flow deviates considerably from simple hydrostatic balance, especially close to the sonic point. In this case, the gas is pushed across the sonic radius primarily by the effect of viscosity which removes angular momentum from the gas rapidly.

Hydrostatic thick disk models have been popular for many years because they have narrow empty funnels along the rotation axis which it is thought could collimate outflows and jets. But the empty funnel is the direct result of two features of these models, namely super-Keplerian rotation which provides the centrifugal force necessary to open up the funnel, and a radial pressure maximum which causes the toroidal morphology. Neither of these features is present in our global solutions with large values of $\alpha$. We therefore suggest that our flows are quite dissimilar to the standard toroidal thick disks. Our flows are likely to be quasi-spherical down to the sonic radius.

Most applications of optically-thin advection-dominated models to real systems (Narayan et al. 1995, 1996, Lasota et al. 1996, Narayan 1996) seem to work best if $\alpha$ is taken to be fairly large, say $\alpha \gtrsim 0.1$. The reason for this is simple. Abramowicz et al. (1995) and Narayan & Yi (1995b) showed that the optically thin branch of solutions exists only below a critical mass accretion rate $\dot{M}_{\rm crit} \sim \alpha^2 \dot{M}_{Edd}$. If $\alpha$ is small, then optically-thin advection-dominated flows are restricted to extremely low mass accretion rates. The luminosities of such flows are then so low that the models are no longer capable of explaining objects we see (Narayan 1996). This empirical evidence therefore argues for a large value of $\alpha$ in naturally occurring advection-dominated flows. By the results of this paper, the accreting gas in such flows very likely assumes a quasi-spherical rather than toroidal morphology.

Does this mean that advection-dominated flows cannot make jets? No, outflows may in fact be likely because the Bernoulli parameter is positive in these flows (see Fig. 7), especially in the regions along the rotation axis (Narayan & Yi 1995a). A bipolar outflow may therefore arise naturally along the rotation axis, though no detailed model has yet been developed. The absence of empty funnels in large $\alpha$ flows, however, implies that any outflow will be pressure-confined right from the point of origin. The pressure-confinement makes our model quite different from other models where the jet is initiated in a low-pressure funnel or corona and is collimated only farther out.

One of the aims of the present work was to compare the exact global solutions with the local self-similar solution obtained by Narayan & Yi (1994) and Spruit et al. (1987). In general, it appears that the local solution does a fairly good job of describing the dynamics of the exact flow. Comparisons between the local and exact solutions are shown in Figs. 1, 2, 6 and 7 to illustrate this point. Also, if the advection-dominated flow attaches to a thin disk on the outside, we find that the transition between the two zones happens quite rapidly. The influence of the outer boundary condition is felt at most over an order of magnitude in radius (Fig. 5). The transonic inner boundary condition has a more pronounced effect because the radial velocity has to increase substantially as the gas approaches the sonic point.

Because viscosity plays an important role in the results, we have tested two different versions of the $\alpha$ prescription for the viscous stress. Most of the calculations are based on the diffusive formulae (2.9), (2.10), but in §3.3 we have repeated some of the calculations using the simpler prescription (3.7), (3.8), and (3.10). As Figs. 8 and 9 show, all the results are reproduced by the second prescription, so that it appears that the precise viscosity prescription is not important. Note, however, that both prescriptions are acausal. The diffusion-type prescription, especially, suffers from infinite propagation speed of viscous signals (e.g. Narayan 1992). Improved theories of viscosity with explicit inclusion of causality are now available (Narayan 1992, Kato & Inagaki 1994, Narayan, Loeb & Kumar 1994, Papaloizou & Szuszkiewicz 1994, Kato 1994, Kato & Yoshizawa 1995) and it would be interesting to study global models with these more physical prescriptions.

The literature on transonic flows around black holes includes many discussions of the nature of the sonic point and its stability characteristics (Liang & Thompson 1980, Matsumoto et al. 1984, Kato, Honma & Matsumoto 1988a, Abramowicz & Kato 1989, Chen & Taam 1993, Kato et al. 1993). No detailed analysis has yet been done for the case of optically-thin, advection-dominated flows, though the results of Chen & Taam (1993) as well as our preliminary work suggest that the sonic points in these flows are always of saddle type and stable. This is in contrast to cooling-dominated flows with the Shakura-Sunyaev type $\alpha$-viscosity, where stable sonic points are restricted to small values of $\alpha \lesssim 0.05 - 0.08$ (Kato, Honma & Matsumoto 1988b). Advection thus seems to enhance the stability of the sonic regions of the flow and to allow physically valid transonic configurations even for values of $\alpha$ as large as 0.3.

Optically-thin advection-dominated disks, however, are still unstable against thermal perturbations if the stabilizing effect of turbulent heat diffusion can be neglected (Kato,

Abramowicz & Chen 1996). This instability is not as serious as one might at first imagine. In the case of cooling-dominated disks, the thermal instability causes a complete breakdown of the equilibrium configuration, posing a serious problem for the very existence of thermally unstable configurations (Pringle, Rees & Pacholczyk 1973, Lightman & Eardley 1974). In advection-dominated disks, on the other hand, the thermal instability does not destroy the flow, since perturbations are propagated into the central object before growing to very large amplitudes. The presence of moderately unstable perturbations in these flows could, in fact, be viewed as a positive feature since it might explain the violent variability observed in black hole X-ray binaries and active galactic nuclei, many of which probably accrete via advection-dominated flows.

Chakrabarti (1990, see also Chakrabarti & Titarchuk 1995) has claimed that at low values of $\alpha$ transonic solutions are replaced by solutions with radial shocks. We do not find the need for such shocks in our solutions. Indeed, we find transonic solutions without shocks for all choices of $\alpha$ in the range 0.001 to 0.3 and $\gamma$ in the range $4/3 - 5/3$.

The differences between our study and Chakrabarti's probably arise from different philosophies on boundary conditions. To check whether the problem is with the outer boundary conditions, we have tried a wide variety of conditions (Fig. 5), spanning the entire range from thin-disk-like conditions (equations 2.13, 2.14) to self-similar conditions (equations 2.21, 2.22). In all cases we find good transonic flows with no shocks near the black hole. Differences in the inner boundary conditions are more likely to be the reason, as we have discussed in §2.2. In our calculations, we determine the sonic radius, $R_s$, and the specific angular momentum accreted by the black hole, $j$, self-consistently. Chakrabarti & Titarchuk (1995) apparently assign an arbitrary value to $j$, a procedure we have trouble understanding.

As we discussed in connection with Fig. 5, large differences in the outer boundary conditions lead to negligibly small changes in $j$. This is because $j$ is determined essentially by boundary conditions on the inside, not the outside (as it would be in the case of an inviscid flow). As an analogy we point out that in the thin accretion disk model $j$ is determined by an inner boundary condition, namely the no-torque condition at the inner edge of the disk (Pringle 1981, Frank et al. 1992). One would not consider selecting $j$ at random, since this would shift the inner edge to a different radius and may well lead to shocks if one tried to match the flow on to a black hole. The reader is referred also to Popham & Narayan (1991, 1992) and Paczyński (1991) for a discussion of the angular momentum eigenvalue in the context of accretion disk boundary layers, where again $j$ is determined by boundary conditions on the inside and cannot be selected arbitrarily. Considering these arguments, it is not surprising that Chakrabarti & Titarchuk find shocks when they choose $j$ at random, but their result may not have any physical significance.

One situation in which a shock can indeed be physical is when the gas is introduced on the outside with an extremely small amount of angular momentum. In this case, the gas would free-fall, as in Bondi accretion, and would shock when it reaches the centrifugal barrier. If the initial angular momentum is low enough, the shock would form close to the black hole, as in Chakrabarti's models, but we do not consider this case very relevant to real flows.

Because advection-dominated flows are very thick in the vertical direction (in fact, they are nearly spherical) it is necessary to question the validity of the height-integration technique used in this paper. It has been shown in Narayan & Yi (1995a) that the one-dimensional height-integrated equations are surprisingly accurate so long as the flow is approximately self-similar. However, our solutions deviate from self-similarity near the boundaries. Moreover, near the inner boundary, the flow accelerates radially and becomes supersonic, which would freeze out vertical motions. These effects are not modeled very accurately in the calculations presented here and it would be of interest to develop improved techniques to handle such complications.

Another issue that needs to be addressed is to combine the treatment of both the dynamics and the cooling in a fully self-consistent way. In the calculations presented here, we either assumed that the cooling is completely negligible at all radii, or we prescribed an arbitrary cooling as a function of radius (equation 3.6 and Fig. 6). In previous papers (e.g. Narayan et al. 1996, Lasota et al. 1996, Narayan 1996) we simplified the dynamics by assuming a self-similar flow, but calculated the cooling in detail including the effects of non-local Comptonization. We also ensured that the cooling was sufficiently small at each radius to permit the advection-dominated branch of solution. In none of the work so far has a detailed cooling calculation been coupled with a self-consistent calculation of the transition radius at which the outer thin disk is transformed into the inner advection-dominated flow. Honma (1996) has calculated some self-consistent models of the transition zone, but with a simplified model of cooling. A more careful analysis of this zone is required.

After the present work was completed, we became aware of the paper by Chen, Abramowicz & Lasota (1996), which is closely related to the present paper.

We are grateful to the second referee for useful criticism which enabled us to improve the presentation. RN thanks Roger Blandford, Piero Madau and Robert Popham for useful discussions. This work was supported in part by NSF grants AST 9423209 (to the Center for Astrophysics) and PHY 9407194 (to the Institute for Theoretical Physics where most of the work was done).

**Figure Captions**

Fig. 1. The solid lines show the variation of the radial velocity $v$ as a function of radius for six advection-dominated solutions. From below, the models correspond to $\alpha = 0.001$, 0.003, 0.01, 0.03, 0.1, and 0.3, respectively. All the models have $\gamma = 1.5$ (corresponding to equipartition between gas and magnetic pressure) and $f = 1$ (fully advection-dominated). The dashed lines show the variation of the sound speed $c_s$ for the same six models, with the lowest curve corresponding to the lowest value of $\alpha$. The lower six dotted lines show the variation of $v(R)$ expected according to the self-similar solution (2.20). The upper dotted line is the self-similar solution (2.21) for $c_s$ in the limit $\alpha^2 \ll 1$.

Fig. 2. The solid curves show the radial variation of the specific angular momentum $l = \Omega R^2$ in the six solutions shown in Fig. 1. The lowest curve corresponds to $\alpha = 0.3$ and the uppermost curve to $\alpha = 0.001$. The dashed line shows the Keplerian specific angular momentum $l_K = \Omega_K R^2$ and the dotted line corresponds to the self-similar solution (2.22).

Fig. 3. Expanded version of Fig. 2, showing the inner region of the flow close to the black hole. Note that the solutions with $\alpha = 0.001$ and 0.003 have $l > l_K$ over a range of radii, whereas the solutions with higher values of $\alpha$ have $l < l_K$ at all radii.

Fig. 4. Variation of gas pressure $p$ with radius for the six solutions in Fig. 1. The curves correspond to decreasing $\alpha$ upwards. Note that the two upper curves have a pressure maximum at $R \sim 4R_g$, exactly where the specific angular momentum crosses the Keplerian value (see Fig. 3). The other solutions with larger values of $\alpha$ do not have pressure maxima.

Fig. 5. The two panels show three solutions corresponding to $\alpha = 0.3$, $\gamma = 1.5$. The solid lines show the solution when we use the boundary conditions, $c_s = 10^{-3}\Omega_K R$, $\Omega = \Omega_K$, at the outer edge of the flow. The dashed lines correspond to the boundary conditions $c_s = (c_s)_{ss}$, $\Omega = \Omega_K$, where the subscript ss refers to the self-similar solution described in §2.3, and the dotted lines correspond to the conditions $c_s = (c_s)_{ss}$, $\Omega = \Omega_{ss}$. Note that the three solutions approach each other at smaller radii, showing that the influence of the outer boundary conditions decays rapidly with decreasing radius.

Fig. 6. Shows a case where $\alpha = 0.1$, $\gamma = 1.5$, and the advection parameter $f$ turns on gradually as a function of $R$ according to the prescription given in equation (3.6). In the upper panel, the solid line shows the global solution for $v$ and the dashed line shows $c_s$. The solid line in the lower panel shows the global specific angular momentum $l$. The dotted lines indicate the self-similar estimates corresponding to the local value of $f(R)$ at each $R$.

Fig. 7. Ther upper panel shows the variation of the Bernoulli parameter $Be$ with radius for the six solutions shown in Fig. 1. The solutions are shown only down to the sonic radius. The dashed line represents the value corresponding to the self-similar solution. The lower panel shows the error parameter $e$ defined in equation (2.4) for the same solutions. This parameter should be near zero (the dashed line) if the approximations used in the paper are to be fully valid.

Fig. 8. Global models obtained with the Shakura-Sunyaev shear stress formula (3.7), instead of the diffusion-type viscosity used in (2.9). As in Fig. 1, the solid and dashed curves show

the radial distributions of the radial velocity $v$ and the sound speed $c_s$, respectively. Values of $\alpha$ in the six models are from below: 0.0005, 0.0015, 0.005, 0.015, 0.05, 0.15. The differences in the values of $\alpha$ shown here compared to those in Fig. 1 are due to the small difference in vertical modeling in the two calculations.

Fig. 9. Angular momentum distributions corresponding to the models shown in Fig. 8. The solid curves show the radial distribution of the specific angular momentum $l = \Omega R^2$. The lowest curve corresponds to $\alpha = 0.05$, and the uppermost curve to $\alpha = 0.0005$. The model with $\alpha = 0.15$ is not shown because $l$ becomes negative. The dashed line corresponds to the Keplerian specific angular momentum $l_K = \Omega_K R^2$.

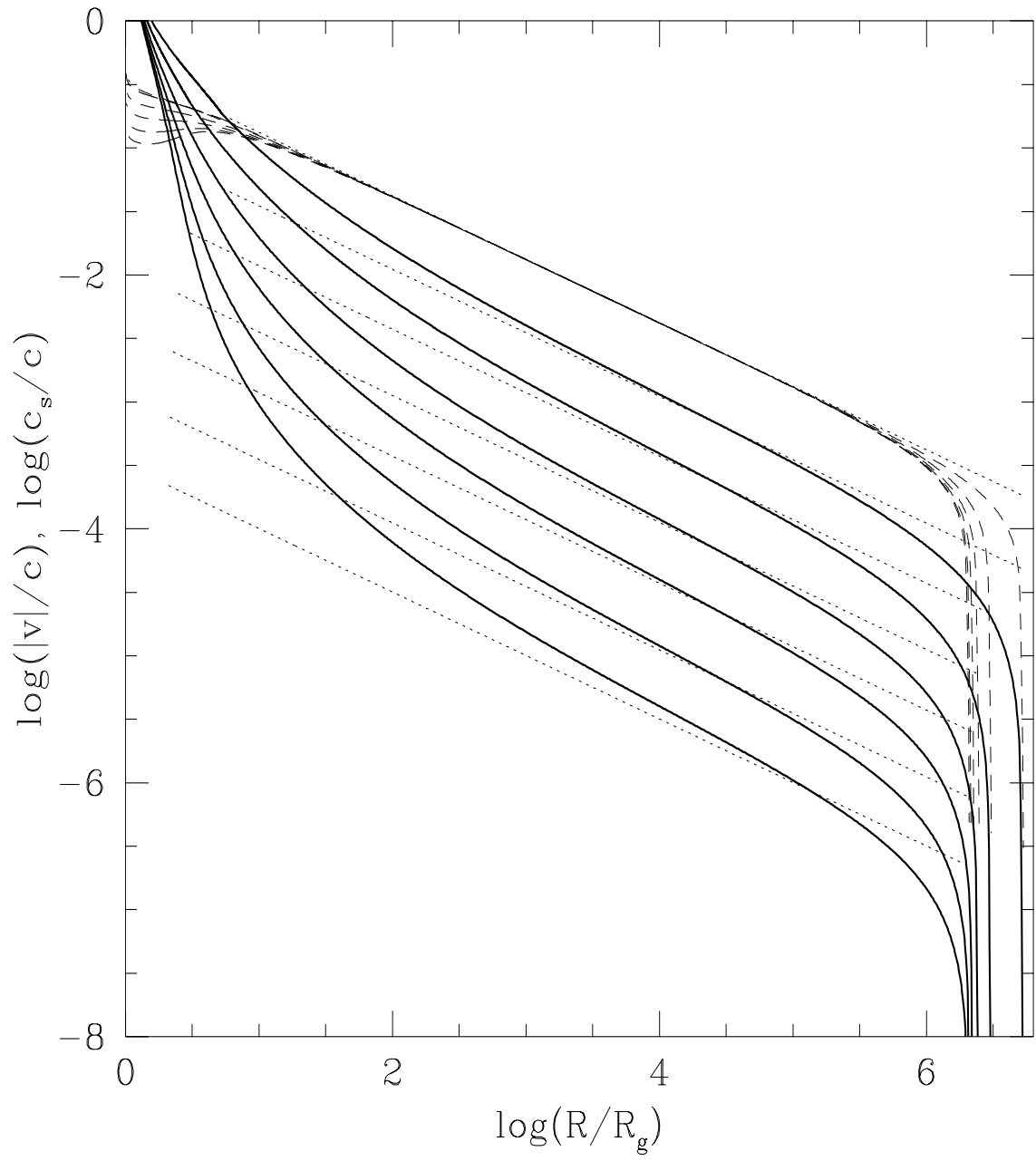

Figure 1

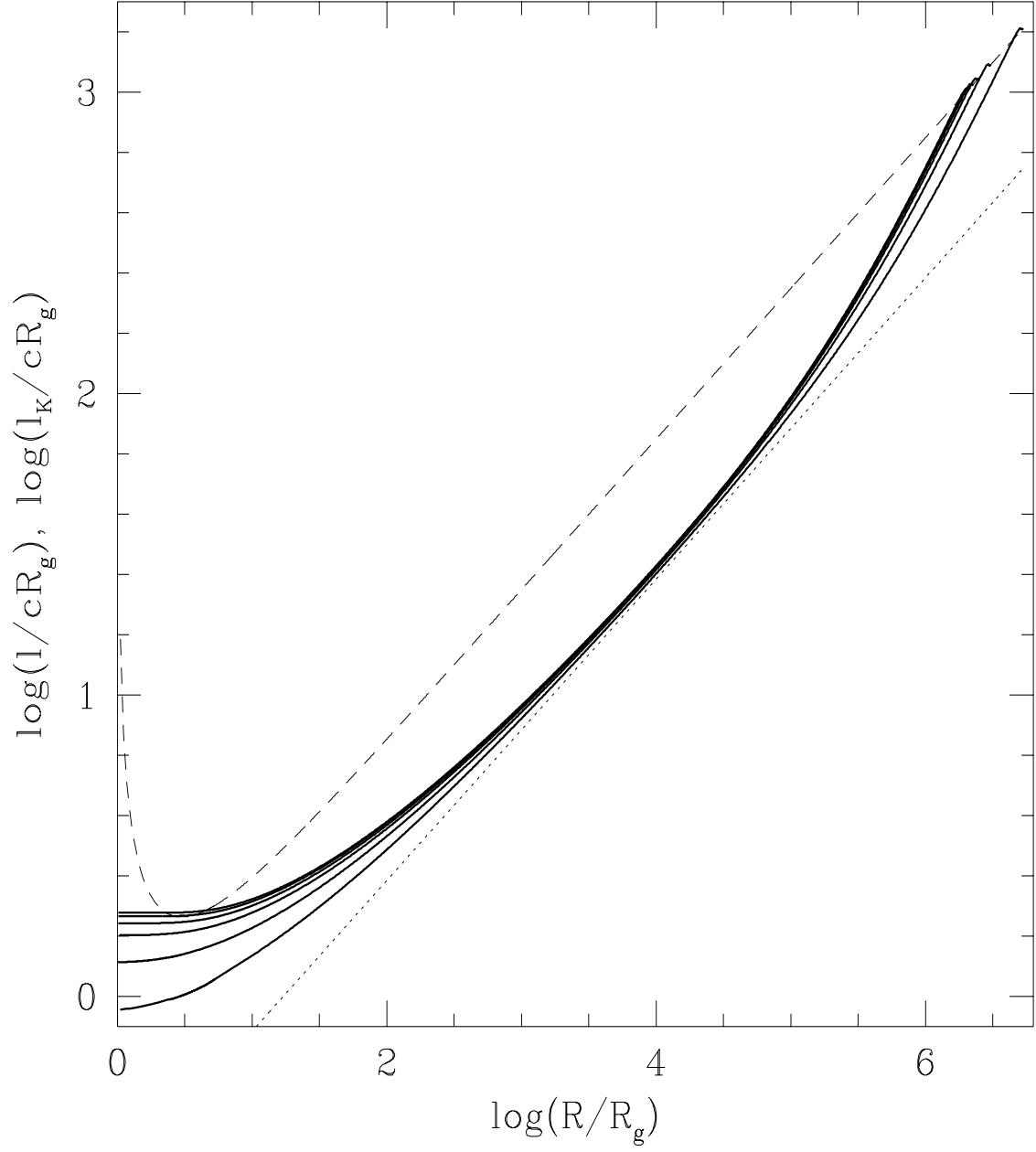

Figure 2

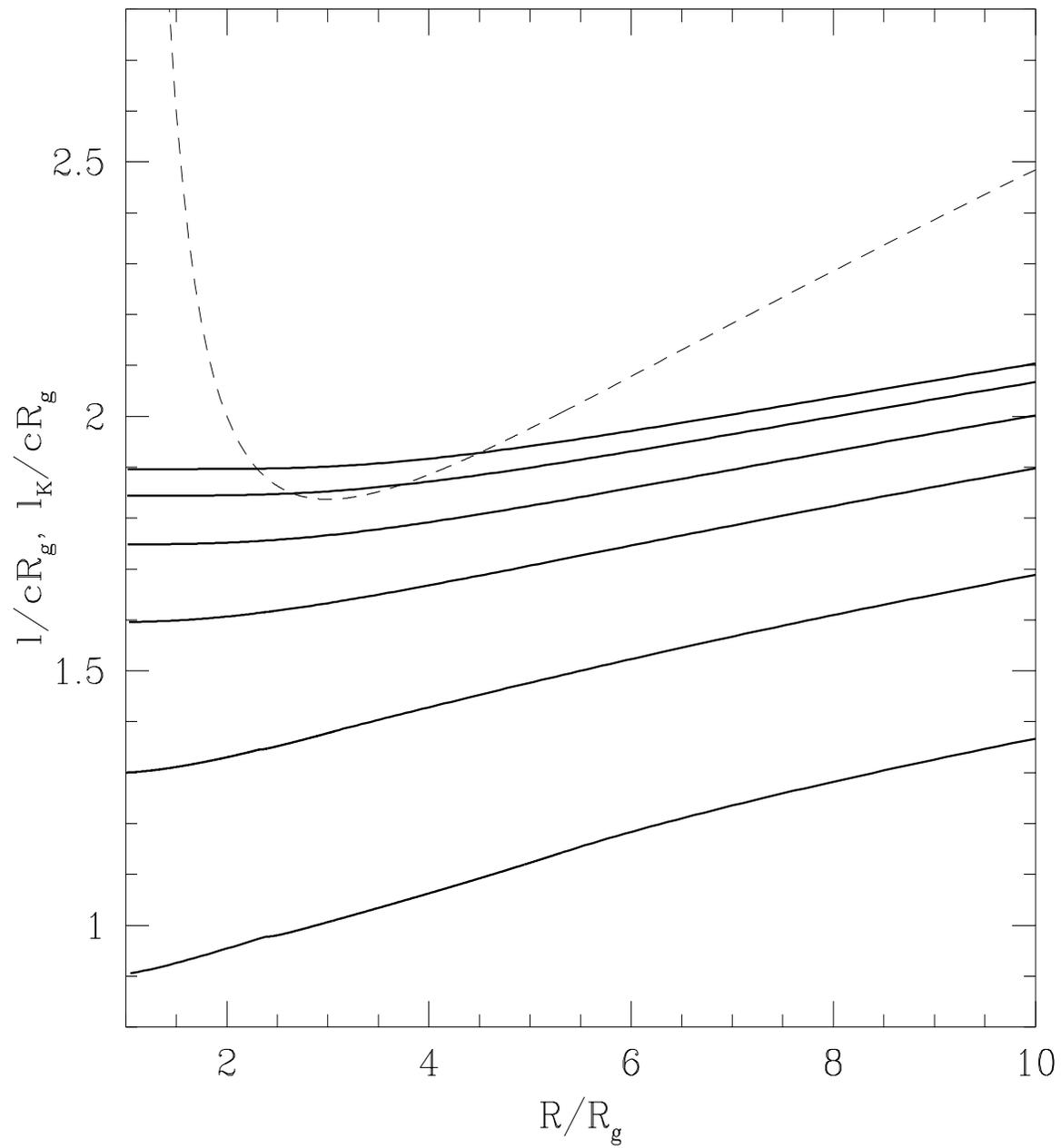

Figure 3

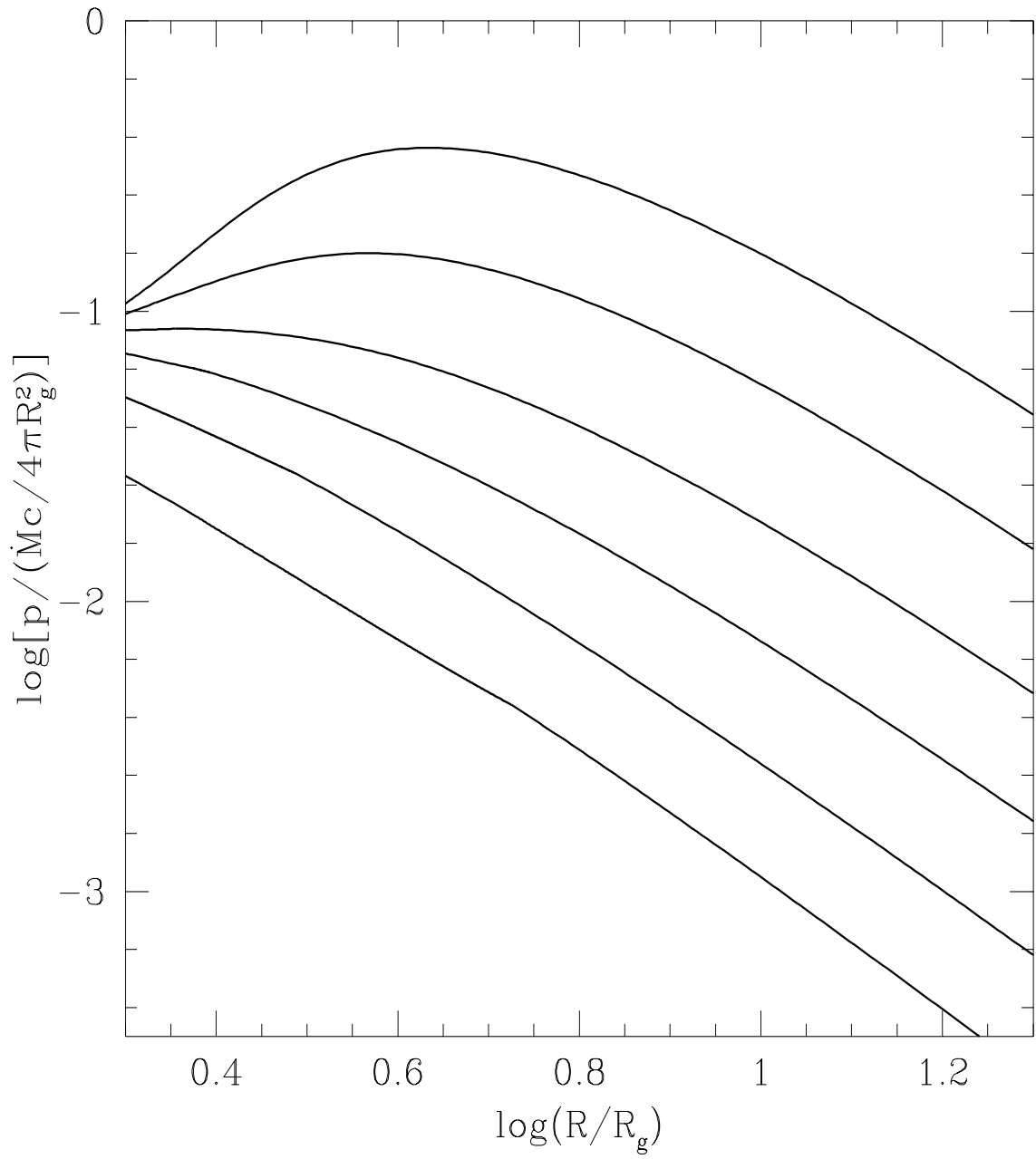

Figure 4

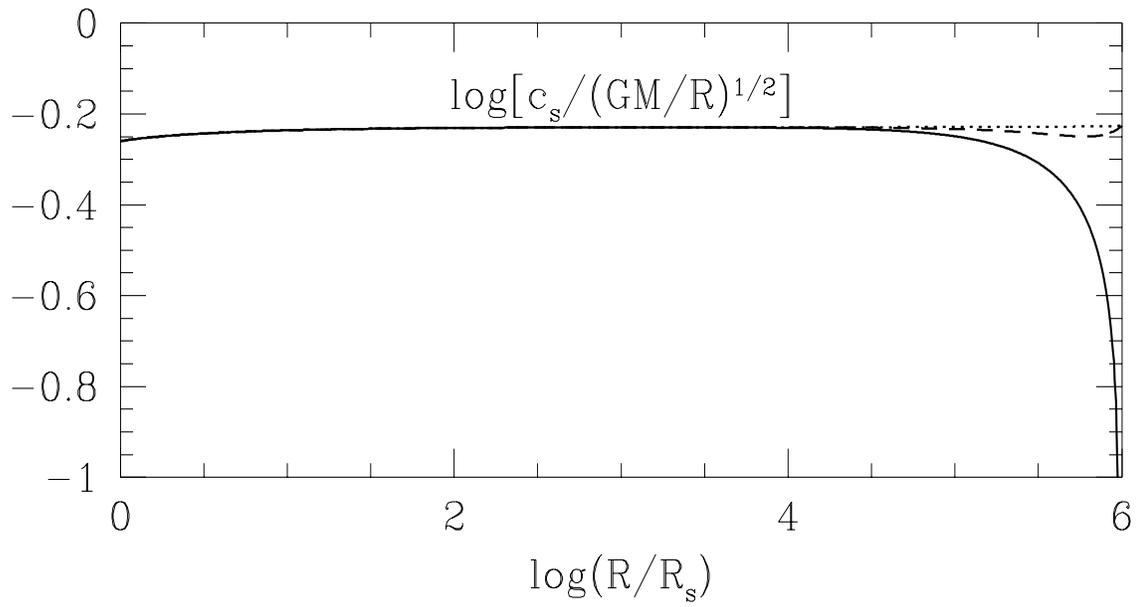

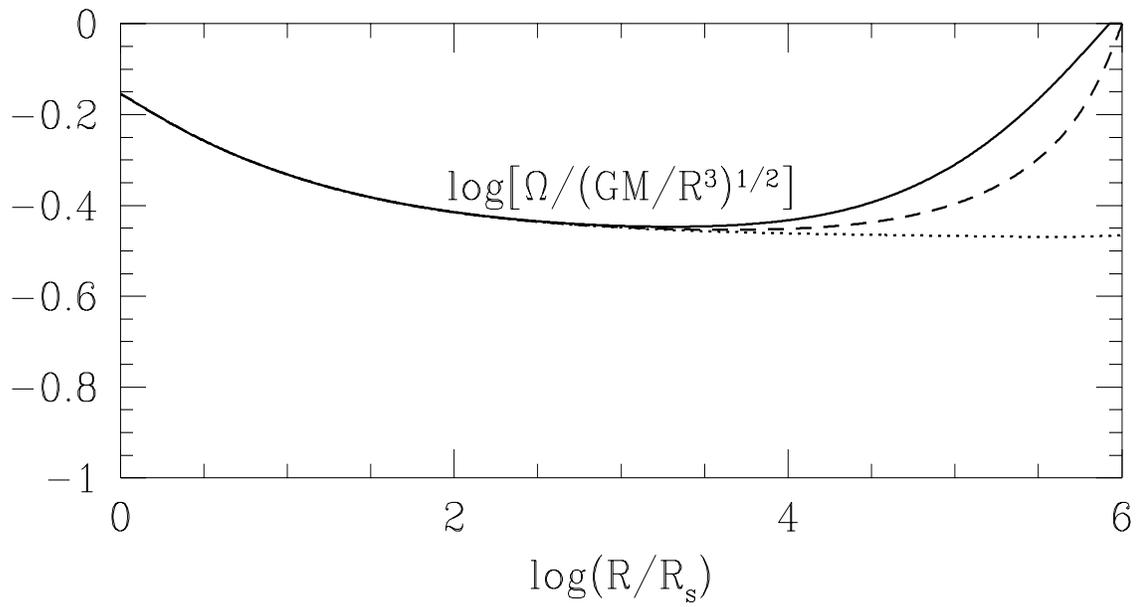

Figure 5

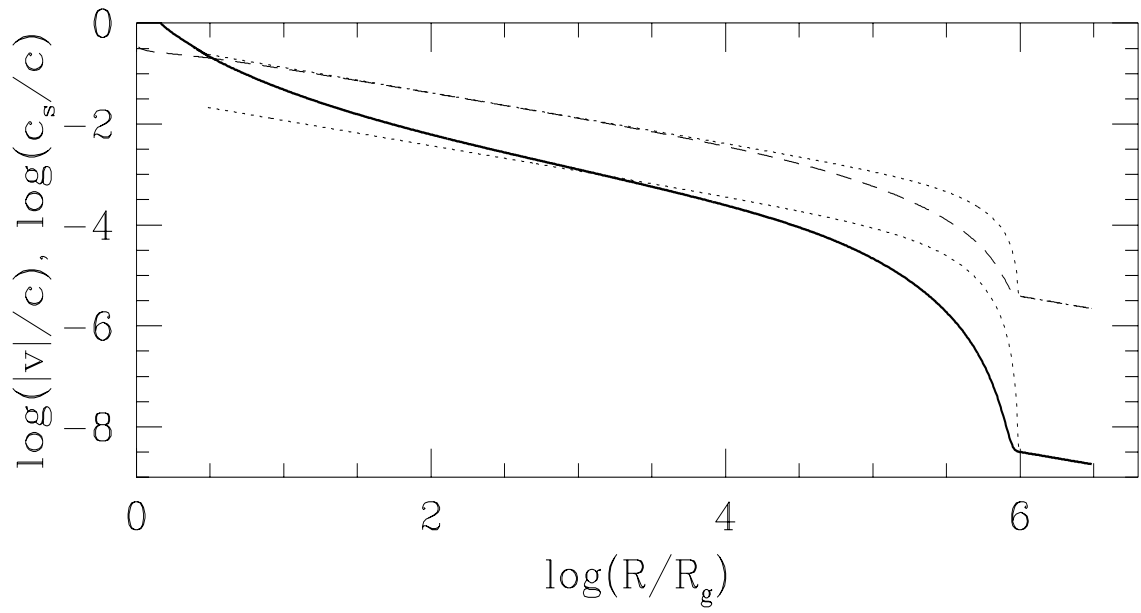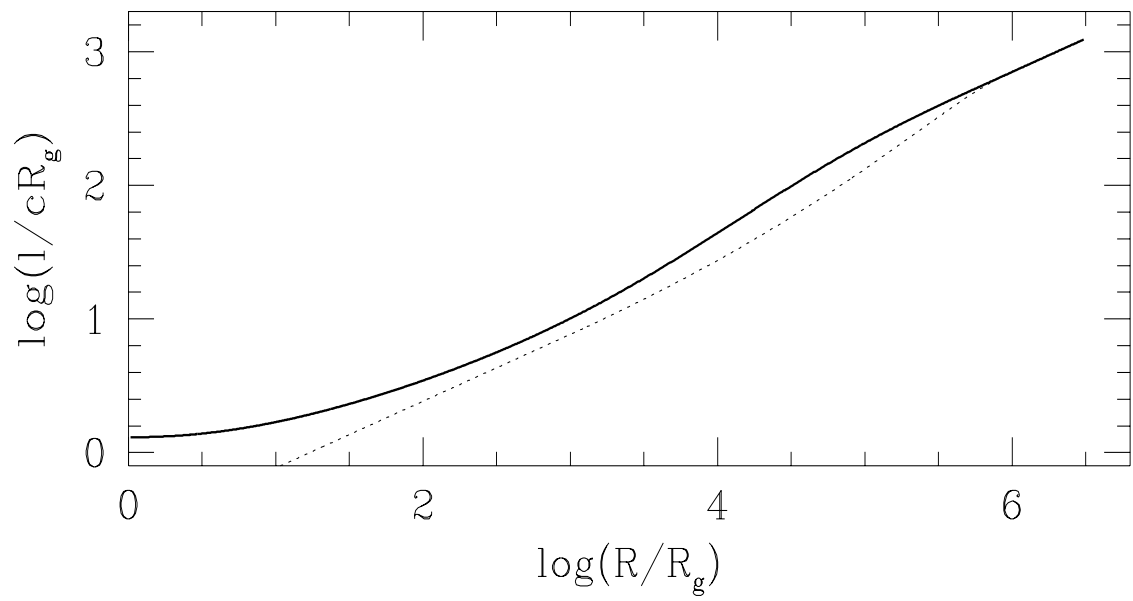

Figure 6

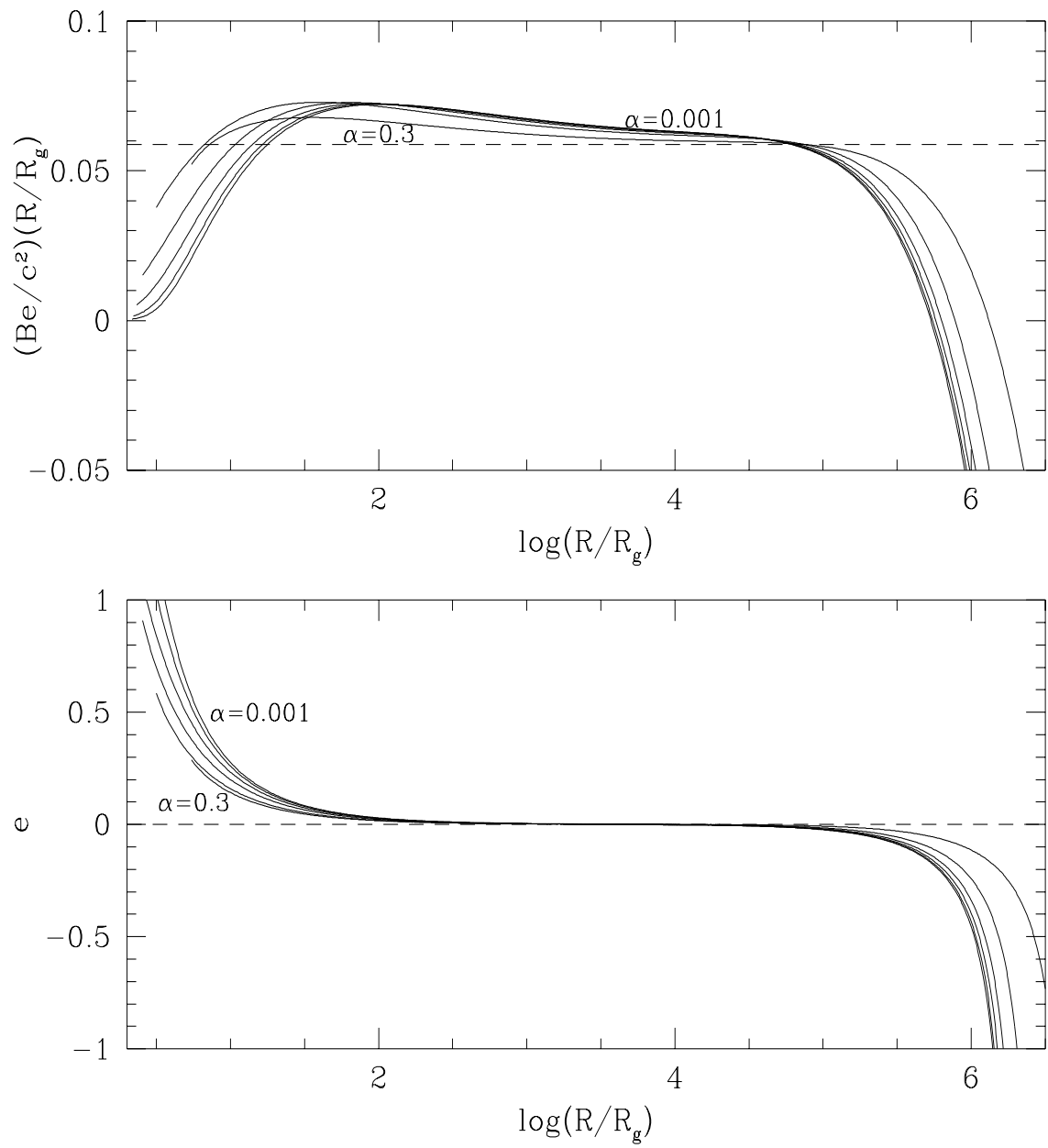

Figure 7

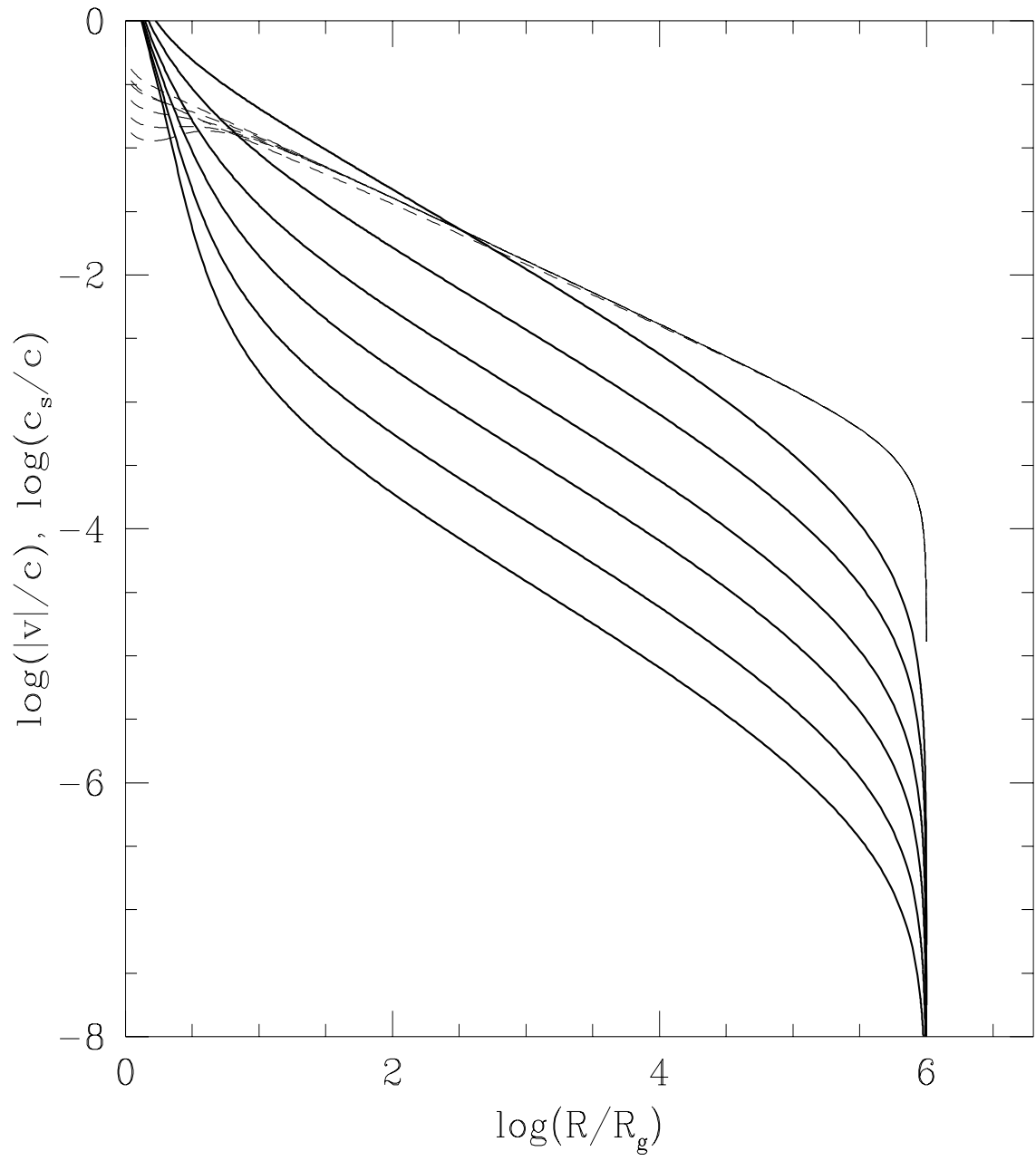

Figure 8

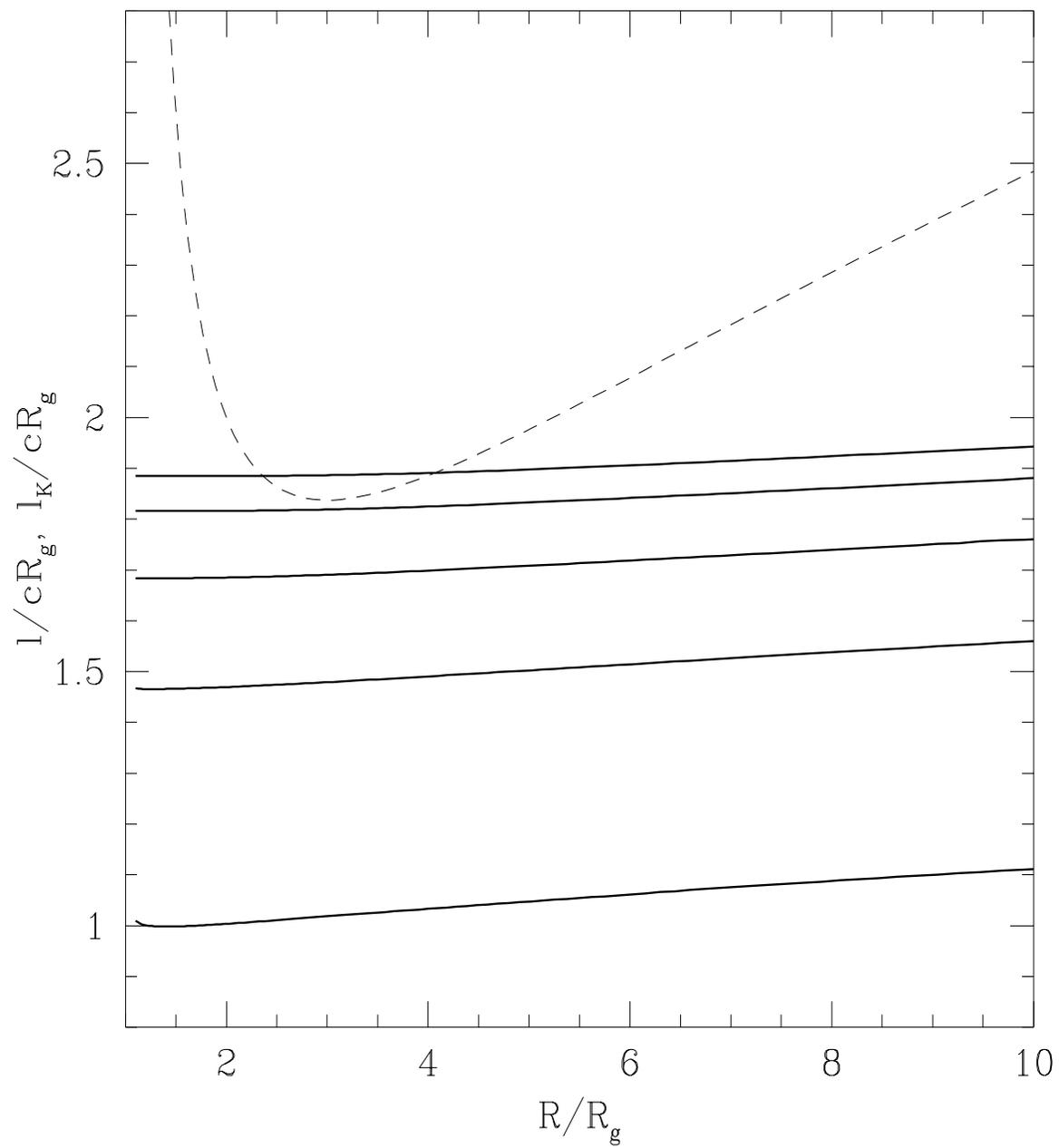

Figure 9